\documentclass[a4paper,12pt]{article}

\usepackage{amsmath}
\usepackage{amssymb}
\usepackage{amsfonts}
\usepackage{amscd}
\usepackage{latexsym}
\usepackage{amsbsy}
\usepackage{bbm}
\usepackage[english]{babel}
\usepackage{graphicx}
\usepackage{psfrag}

\begin{document}

\begin{center}

\vspace{.7cm}
{\LARGE {\bf From D3-Branes to Lifshitz\\  \vspace{0.2cm}  Space-Times}} \\

\vspace{1.2cm}

{\large Wissam Chemissany$^a$ and Jelle Hartong$^b${}\\
\vspace{1cm}
 $^a$ {\small\slshape Instituut voor Theoretische Fysica, Katholieke Universiteit Leuven,\\ Celestijnenlaan 200D, B-3001 Leuven, Belgium }\\\vspace{0.2cm}

$^b${\small\slshape Albert Einstein Center for Fundamental Physics,\\
 Institute for Theoretical Physics,\\
 University of Bern,\\
 Sidlerstrasse 5, CH-3012 Bern,
 Switzerland}}\\

 {\small\slshape Email: wissam@itf.fys.kuleuven.be, hartong@itp.unibe.ch}
\vspace{1cm}

 {\bf Abstract} \end{center} { }

We present a simple embedding of a $z=2$ Lifshitz space-time into type IIB supergravity. This is obtained by considering a stack of D3-branes in type IIB supergravity and deforming the world-volume by a plane wave. The plane wave is sourced by the type IIB axion. The superposition of the plane wave and the D3-branes is 1/4 BPS. The near horizon geometry of this configuration is a 5-dimensional $z=0$ Schr\"odinger space-time times a 5-sphere. This geometry is also 1/4 BPS. Upon compactification along the direction in which the wave is traveling the 5-dimensional $z=0$ Schr\"odinger space-time reduces to a 4-dimensional $z=2$ Lifshitz space-time. The compactification is such that the circle is small for weakly coupled type IIB string theory. This reduction breaks the supersymmetries. Further, we propose a general method to construct analytic $z=2$ Lifshitz black brane solutions. The method is based on deforming AdS$_5$ black strings by an axion wave and reducing to 4-dimensions. We illustrate this method with an example.

\newpage

\tableofcontents

\section{Introduction}

The Lifshitz symmetry group is one of the possible symmetry groups that one encounters at quantum critical points in condensed matter systems. Typically such systems are strongly coupled and the idea is more or less that, much like in AdS/CFT, the universal behavior of some strongly coupled field theory near a quantum critical point can be studied holographically by means of a suitably chosen gravitational background such as an asymptotically Lifshitz space-time. The subject of Lifshitz holography was initiated by \cite{Kachru:2008yh}. For more details about the basic holographic properties of Lifshitz space-times see \cite{Taylor:2008tg}.

Recently a lot of progress has been made regarding the problem of embedding Lifshitz space-times into string theory \cite{Hartnoll:2009ns,Balasubramanian:2010uk,Donos:2010tu,Gregory:2010gx,Cassani:2011sv,Halmagyi:2011xh,Narayan:2011az}. We present a particularly simple way of embedding a 4-dimensional $z=2$ Lifshitz space-time into type IIB string theory. Our construction is inspired by \cite{Cassani:2011sv} (and also appears in \cite{Donos:2010tu}) and works as follows. Take the bosonic part of type IIB supergravity and truncate the 2-forms. Reduce this theory over a 5-sphere of constant radius and subsequently perform a Scherk--Schwarz reduction in which the axion shift symmetry is gauged. The resulting 4-dimensional theory admits $z=2$ Lifshitz solutions. In this construction the dilaton is a free constant and the axion is a linear function of the compactification circle.

This way of embedding Lifshitz into type IIB supergravity immediately points towards a D3-brane interpretation. We simply take an extremal D3-brane and add to it a solution that is sourced by the axion. This latter solution turns out to be a 1/2 BPS plane wave. The intersection of the plane wave with the extremal D3-brane is 1/4 BPS and its near horizon geometry is a 5-dimensional $z=0$ Schr\"odinger space-time. Such a space-time can be reduced to a 4-dimensional $z=2$ Lifshitz space-time, see for example \cite{Balasubramanian:2010uk}.

Once we have established such a D3-brane/axion wave interpretation of the Lifshitz space-time we address the question of finding black deformations. We propose a general method to construct analytic $z=2$ Lifshitz black brane solutions and illustrate it with a simple example.

\section{D3-branes and plane waves}\label{sec:D3branesandwaves}

Consider type IIB supergravity and truncate away the fermions and the 2-forms. The equations of motion of the resulting theory are (we use the conventions of \cite{Schwarz:1983qr})
\begin{align}
& R_{AB} = P_AP_B^*+P_BP_A^*+\frac{1}{6}F_{C_1\ldots C_4A}F^{C_1\ldots C_4}{}_B\,,\\
& \mathcal{D}_AP^A=\partial_AP^A+\Gamma_{AB}^AP^B-2iQ_AP^A=0\,,\\
& \star F_5=F_5\,,\qquad dF_5=0\,.
\end{align}
By capital Latin indices we denote 10-dimensional indices. We denote by $\mathcal{D}_A$ the covariant derivative that contains besides the Levi--Civit\`a connection (and later when we look at Killing spinors the Lorentz connection) also a $U(1)$ connection $Q_A$ which is the pull-back of the K\"ahler connection 1-form of the scalar coset space $SL(2,\mathbb{R})/SO(2)$. We choose the usual axion-dilaton parametrization of the scalar coset manifold in which $P_A$ and $Q_A$ are
\begin{eqnarray}
P_A & = & \frac{1}{2}\partial_A\phi+\frac{i}{2}e^\phi\partial_A\chi\,,\\
Q_A & = & \frac{1}{2}e^\phi\partial_A\chi\,.
\end{eqnarray}

The solution that we will be interested in describes a plane wave traveling along the world-volume of a stack of D3-branes and is given by
\begin{align}
& ds^2 = H^{-1/2}\left(2dtdu+Gdu^2+dx^2+dy^2\right)+H^{1/2}\left(dR^2+R^2d\Omega_5^2\right)\,,\label{eq:D3braneplanewave1}\\
& F_{\hat\mu_1\ldots\hat\mu_5} = l^4R^{-5}H^{-5/4}\epsilon_{\hat\mu_1\ldots\hat\mu_5}\,,\qquad F_{a_1\ldots a_5}=l^4\omega_{a_1\ldots a_5}\,,\\
& \chi=ku\,,\\
& \phi=\phi_0\,,\label{eq:D3braneplanewave4} 
\end{align}
in which the functions $H$ and $G$ are 
\begin{eqnarray}
H & = & 1+\frac{l^4}{R^4}\,,\\
G & = & \frac{1}{4}e^{2\phi_0}k^2\left(\frac{l^4}{R^2}-\frac{R^2}{3}\right)\,.\label{eq:functionG}
\end{eqnarray}
The parameters $l$ and $k$, relate to the D3-brane charge and the wave momentum, respectively. The 5-sphere is parametrized by the coordinates $x^a=(\theta_1,\ldots,\theta_4,\varphi)$ and $\epsilon_{\hat\mu_1\ldots\hat\mu_5}$ and $\omega_{a_1\ldots a_5}$ denote the volume forms on the 5-dimensional non-compact space-time parametrized by $x^{\hat\mu}=(t,u,x,y,R)$ and the 5-sphere, respectively. The dilaton assumes a constant value $\phi_0$ and from now on we will write $e^{\phi_0}$ as $g_s$, the type IIB string coupling. When we set $k$ equal to zero we recover the well-known extremal D3-brane solution and when we set $l$ equal to zero we obtain a plane wave space-time. We note that it is not possible to transform away the string coupling $g_s$ from the solution by a rescaling of the coordinates $t$ and $u$. This can be used to remove $g_s$ from the metric but then it will reappear in the expression for the axion. The solution \eqref{eq:D3braneplanewave1} to \eqref{eq:D3braneplanewave4} is a special case of the solutions presented in \cite{Donos:2010tu} corresponding to constant dilaton and zero NSNS and RR 3-form field strengths.

The near horizon geometry is obtained by considering the solution near $R=0$. Defining $r=l^2R^{-1}$ we find
\begin{align}
& ds^2 = l^2\left[\frac{1}{r^2}\left(2dtdu+\frac{1}{4}g_s^2k^2r^2du^2+dx^2+dy^2+dr^2\right)+d\Omega_5^2\right]\,,\label{eq:nearhorizonmetric}\\
& F_{\hat\mu_1\ldots\hat\mu_5} = l^4\epsilon_{\hat\mu_1\ldots\hat\mu_5}\,,\qquad F_{a_1\ldots a_5}=l^4\omega_{a_1\ldots a_5}\,,\label{eq:nearhorizon5form}\\
& \chi=ku\,,\\
& e^\phi=e^{\phi_0}=g_s\,.\label{eq:nearhorizonphi}
\end{align}
The metric describes the direct product of a 5-dimensional $z=0$ Schr\"odinger space-time and the 5-sphere. This solution of type IIB supergravity has also been studied in \cite{Narayan:2011az}. In the near horizon geometry when the parameter $k$ is nonzero it can be rescaled away by rescaling $u$ and $t$.

The symmetries preserved by this solution are translations in $t,x,y$, rotation of $x$ and $y$ together with the following three transformations
\begin{align}
&r\rightarrow\lambda r\,,\qquad \vec x\rightarrow\lambda\vec x\,,\qquad t\rightarrow\lambda^2 t\,,\\
&\vec x\rightarrow\vec x-\vec v u\,,\qquad t\rightarrow\vec v\cdot\vec x-\frac{1}{2}\vec v^2 u\,,
\end{align}
where $\vec x=(x,y)$ and $\vec v$ is a constant vector. The first of these transformations is a dilatation symmetry and the other two are boost transformations with respect to ``$u$-velocities''. The translations in $u$, that are usually considered part of the $z=0$ Schr\"odinger algebra, are broken by the axion, though the violation is mild in the sense that the axion transforms into itself up to a symmetry of type IIB supergravity. 

In terms of Killing vectors the symmetries of the 5-dimensional $z=0$ Schr\"odinger space-time are
\begin{eqnarray}
H & = & \frac{\partial}{\partial t}\,,\\
\vec P & = & \frac{\partial}{\partial\vec x}\,,\\
\vec V & = & -\vec x\frac{\partial}{\partial t}+u\frac{\partial}{\partial\vec x}\,,\\
M & = & x\frac{\partial}{\partial y}-y\frac{\partial}{\partial x}\,,\\
D & = & 2t\frac{\partial}{\partial t}+\vec x\cdot\frac{\partial}{\partial\vec x}+r\frac{\partial}{\partial r}\,,\\
N & = & \frac{\partial}{\partial u}\,.
\end{eqnarray}
This algebra contains the Carroll algebra, made out of $H,\vec P,\vec V$ and $M$ as a subalgebra. The Carroll algebra is the contraction of the Poincar\'e algebra in which the speed of light is sent to zero.

For arbitrary values of $z$ the metric of the Schr\"odinger space-time takes the form given above with the $du^2$ term replaced by $r^{-2z}du^2$. Changing a bit the parameters this metric can be written as
\begin{equation}\label{eq:Schmetric}
 ds^2 = \frac{1}{r^2}\left(2dtdu+dx^2+dy^2+dr^2\right)\pm r^{-2z}du^2\,,
\end{equation}
putting temporarily $l=1$. Schr\"odinger space-times come in two different forms that differ merely by a sign in front of the $r^{-2z}du^2$ term in the metric, but that have the same Lie algebra of Killing vectors. They are nonetheless different space-times, i.e. not related by a diffeomorphism. The choice of sign has important consequences for the singular structure of the space-time. From a combination of a tidal force computation and a study of the geodesic (in)completeness of the metric \eqref{eq:Schmetric} we conclude that with $0\le z<1$ and for the minus sign in front of the $du^2$ term the metric is geodesically complete and the space-time is free of tidal force singularities (this follows from the results\footnote{The sign choice in the metric does not strongly affect the calculation of the tidal forces. For example, if we take a congruence of timelike geodesics with tangent $u^\mu$ and with $\dot x=\dot y=0$ then $R_{\underline{x}\mu\nu\underline{x}}u^\mu u^\nu=-1\pm P_t^2(z-1)r^{4-2z}$ in which $P_t$ is a constant of the geodesic motion ($\dot u=P_t r^2$). The vielbeins $e^{\underline{x}}_\mu=r^{-1}\delta_\mu^x$ and $u_\mu$ are part of an orthonormal frame that is parallely propagated along these geodesics. Underlined indices are tangent space indices. We thus observe that the tidal forces diverge as $r$ goes to infinity. Therefore the question is can timelike geodesics get to $r=\infty$ in finite proper time or not and it is the answer to this question that strongly depends on the sign choice in the metric.} of \cite{Blau:2009gd}) whereas for the plus sign the metric is geodesically incomplete \cite{Costa:2010cn} and the tidal forces become infinite as $r\rightarrow\infty$. We conclude that the $z=0$ Schr\"odinger space-time in \eqref{eq:nearhorizonmetric} possesses tidal force singularities.

The near horizon solution \eqref{eq:nearhorizonmetric} to \eqref{eq:nearhorizonphi} solves the equations of motion of the following 5-dimensional action
\begin{equation}\label{eq:5Daction}
S=\frac{1}{16\pi G_N^{(5)}}\int d^5x\sqrt{-\hat g}\left(\hat R+\frac{12}{l^2}-\frac{1}{2}\partial_{\hat\mu}\hat\phi\partial^{\hat\mu}\hat\phi-\frac{1}{2}e^{2\hat\phi}\partial_{\hat\mu}\hat\chi\partial^{\hat\mu}\hat\chi\right)\,,
\end{equation}
where hatted fields refer to 5-dimensional fields. The 5-form has been dualized to a cosmological constant term. This action, as is well-known, follows by performing a Freund--Rubin compactification of type IIB supergravity over the 5-sphere.

If we compactify $u\sim u+2\pi L$, so that $u$ parametrizes a spacelike circle, and reduce to 4 dimensions, we obtain a $z=2$ Lifshitz space-time. To make this more explicit we write the $z=0$ Schr\"odinger metric in the form of a Kaluza--Klein reduction Ansatz
\begin{eqnarray}
ds^2 & = & l^2\left[-\frac{4}{k^2g_s^2}\frac{dt^2}{r^4}+\frac{1}{r^2}\left(dx^2+dy^2+dr^2\right)\right.\nonumber\\
&&\left.+\frac{k^2g_s^{2}}{4}\left(du+\frac{4}{k^2g_s^2}\frac{dt}{r^2}\right)^2\right]\,.
\end{eqnarray}

Asymptotically Lifshitz solutions can be studied as solutions of a 4-dimensional action that follows from the above 5-dimensional action \eqref{eq:5Daction} by Scherk--Schwarz reduction in which the axion shift symmetry has been gauged (see also \cite{Cassani:2011sv}). We make the following reduction Ansatz
\begin{eqnarray}
 d\hat s^2 & = & e^{-\Phi}ds^2+e^{2\Phi}\left(du+A\right)^2\,,\\
 \hat\chi & = & \chi+ku\,,\\
 \hat\phi & = & \phi\,,
\end{eqnarray}
where the unhatted fields\footnote{To avoid heavy notation we use the same symbols to denote 10- and 4-dimensional fields. We hope this does not lead to any confusion.} do not depend on $u$. The resulting 4-dimensional action is
\begin{eqnarray}
 S & = & \frac{1}{16\pi G_N^{(4)}}\int d^4x\sqrt{-g}\left(R-\frac{1}{2}\left(\partial\phi\right)^2-\frac{1}{2}e^{2\phi}\left(D\chi\right)^2-\frac{3}{2}\left(\partial\Phi\right)^2\right.\nonumber\\
&&\left.-\frac{1}{4}e^{3\Phi}F^2-V\right)\,,\label{eq:IIB4Daction}
\end{eqnarray}
where
\begin{eqnarray}
 D\chi & = & d\chi-kA\,,\\
 F & = & dA\,,\\
 V & = & \frac{k^2}{2}e^{-3\Phi+2\phi}-\frac{12}{l^2}e^{-\Phi}\,.
\end{eqnarray}
The equations of motion coming from this action admit the following Lifshitz invariant solution
\begin{eqnarray}
 ds^2 & = & l^3\frac{kg_s}{2}\left[-\frac{4}{k^2g_s^2}\frac{1}{r^4}dt^2+\frac{1}{r^2}dr^2+\frac{1}{r^2}\left(dx^2+dy^2\right)\right]\,,\label{eq:Lifsolution1}\\
 A & = & \frac{4}{k^2g_s^{2}}\frac{1}{r^2}dt\,,\label{eq:Lifsolution2}\\
 e^{\Phi} & = & l\frac{kg_s}{2}\,,\label{eq:Lifsolution3}\\
 e^\phi & = & g_s\,,\label{eq:Lifsolution4}\\
 \chi & = & \chi_0\,,\label{eq:Lifsolution5}
\end{eqnarray}
where $\chi_0$ is an arbitrary constant. The type IIB coupling constant $g_s$ is a free parameter. For $k=0$ this metric makes no sense since for that case the reduction cannot be done along $u$, it being a lightlike circle for $k=0$.

To get this 4-dimensional model we first reduced over the 5-sphere and then along a circle. We could have also done this the other way around. The Scherk--Schwarz reduction of the IIB axion shift symmetry along a circle gives rise to a massive deformation of 9-dimensional maximal supergravity \cite{Meessen:1998qm,Bergshoeff:2002nv} and 4-dimensional Lifshitz times the 5-sphere is a solution of that theory.

The supergravity approximation, i.e. small curvature and weak string coupling requires
\begin{equation}
\frac{l}{l_s}\gg 1\,,\qquad g_s\ll1\,.
\end{equation}
The radius of the circle over which we compactify from 5 to 4 dimensions is given by (in units of string length)
\begin{equation}
\frac{1}{l_s}\int_0^{2\pi L}du \sqrt{g_{uu}}=\frac{l}{l_s}\pi kL g_s\,.
\end{equation}
Using that in type IIB string theory the axion shift symmetry is broken to a symmetry under integer shifts (which here follows from single-valuedness of the axion wavefunction along $u$) we conclude that $2\pi Lk\in\mathbb{Z}$. Hence the circle is of small radius, i.e. of the order of the string length if $lg_s/l_s$ is of order unity. Assuming weak curvature implies that then the type IIB string coupling is weak.

We end this section with a few remarks about the solution \eqref{eq:D3braneplanewave1} to \eqref{eq:D3braneplanewave4} and Kaluza--Klein reduction of Schr\"odinger space-times. In the expression for $G$, equation \eqref{eq:functionG}, we have set two integration constants equal to zero. The most general solution for $G$ is
\begin{equation}
 G= \frac{1}{4}e^{2\phi_0}k^2\left(\frac{l^4}{R^2}-\frac{R^2}{3}\right)+c_1+\frac{c_2}{R^{4}}\,,
\end{equation}
where the constant $c_1$ can be put to zero by the coordinate transformation $t\rightarrow t-c_1u/2$. The constant $c_2$ has been put to zero by hand in \eqref{eq:functionG}. For $k=0$ and $c_2\neq 0$ the solution for $G$ can be obtained from the solution generating technique of Garfinkle and Vachaspati \cite{Garfinkle:1990jq}. It is interesting to note that for $k=0$ and $c_2\neq 0$ we have in the near horizon limit, i.e. for $H=l^4R^{-4}$, what might be called a 5-dimensional $z=-1$ Schr\"odinger space-time. For $c_2>0$ this can be ``Kaluza--Klein reduced'' to a 4-dimensional $z=3$ Lifshitz space-time \cite{Singh:2010zs}. In \cite{Singh:2010zs} the 5-dimensional $z=-1$ Schr\"odinger space-time is obtained via a scaling limit of the near horizon limit of a boosted black D3-brane.

In general if we take the Schr\"odinger metric \eqref{eq:Schmetric} with the plus sign and with $u$ periodically identified, then the ``Kaluza--Klein reduction'' along the spacelike circle\footnote{By saying that $u$ parametrizes a spacelike direction and thus that the compactification circle is spacelike we mean that the tangent of a curve moving only in the $u$-direction is spacelike when measured with respect to the bulk metric. We call $u$ spacelike even when $u$ parametrizes a null direction with respect to the boundary metric.} parametrized by $u$ gives rise to a Lifshitz space-time with dynamical exponent $2-z$. We write Kaluza--Klein reduction in quotation marks because for $z\neq 0$ the 4-dimensional metric in Einstein frame is only conformal to a Lifshitz space-time with dynamical exponent $2-z$ and further because for $z<0$ the physical size of the compactification circle goes to zero near the boundary ($r=0$ or $R=\infty$) while for $z>0$ the physical size goes to zero for $r=\infty$ (or $R=0$). Hence for $z\neq 0$ one would expect that the periodic identification of $u$ will violate the supergravity approximation as light string winding modes cannot be ignored. The case $z=0$ is special because the 4-dimensional space-time in Einstein frame is a $z=2$ Lifshitz space-time and further the physical size of the compactification circle is independent of the radial coordinate. This agrees with the fact that for $z=0$ the Killing vectors $D$ (dilatations) and $N$ (translations in $u$) commute. It has been pointed out in \cite{Costa:2010cn} that for $z<1$ the periodic identification of the $u$ coordinate introduces a null identification in the AdS$_5$ boundary metric and thus corresponds to some DLCQ of the boundary field theory. This is because the $du^2$ term in the metric for $z<1$ is subleading to the AdS part.

\section{Supersymmetry}\label{sec:susy}

The type IIB Killing spinor equations are 
\begin{eqnarray}
\delta\lambda & = & iP_A\gamma^A\epsilon_C=0\,,\\
\delta\psi_A & = & \mathcal{D}_A\epsilon+\frac{i}{480}\gamma^{C_1\ldots C_5}\gamma_A F_{C_1\ldots C_5}\epsilon=0\,,
\end{eqnarray}
in which
\begin{equation}
\mathcal{D}_A\epsilon = \left(\partial_A+\frac{1}{4}\omega_A{}^{\underline{A}\underline{B}}\gamma_{\underline{A}\underline{B}}-\frac{i}{2}Q_A\right)\epsilon\,.
\end{equation}
Underlined indices are tangent space indices. The spinors $\lambda$, $\psi_\mu$ and $\epsilon$ are complex combinations of the two type IIB Majorana--Weyl spinors, e.g. $\epsilon=\epsilon_1+i\epsilon_2$ with $\epsilon_1$ and $\epsilon_2$ Majorana--Weyl. By $\epsilon_C$ we denote the charge conjugate spinor $\epsilon_C=\epsilon_1-i\epsilon_2$ and similarly for $\lambda$ and $\psi_\mu$.

We will show that the D3-brane/plane wave intersection as well as its near horizon geometry are 1/4 BPS solutions. The integrability condition for the gravitino equation is
\begin{eqnarray}
[\mathcal{D}_A,\mathcal{D}_B]\epsilon & = & \left(\frac{1}{4}R_{AB}{}^{\underline{A}\underline{B}}\gamma_{\underline{A}\underline{B}}+\frac{1}{2}\left(P_AP^*_B-P_BP^*_A\right)\right)\epsilon\nonumber\\
& = & \left(\nabla_AS_B-\nabla_BS_A-\left(S_AS_B-S_BS_A\right)\right)\epsilon\,,
\end{eqnarray}
where $R_{AB}{}^{\underline{A}\underline{B}}$ is the curvature 2-form. By $S_A$ we denote
\begin{equation}
S_A=-\frac{i}{480}\gamma^{C_1\ldots C_5}\gamma_AF_{C_1\ldots C_5}\,.
\end{equation}
The integrability condition for the dilatino equation can be obtained by acting with $\mathcal{D}_A$ on the right hand side of $\delta\lambda$ and using the gravitino equation to rewrite $\mathcal{D}_A\epsilon$. Using that $\lambda$ has $U(1)$ weight 3/2 and $P$ has $U(1)$ weight $2$ we find (we actually used the charge conjugate dilatino equation)
\begin{equation}
\gamma^B\left(\mathcal{D}_AP_B^*-P_B^*S_A\right)\epsilon=0\,.
\end{equation} 
Substituting the complete solution \eqref{eq:D3braneplanewave1} to \eqref{eq:D3braneplanewave4} into these two integrability conditions leads to the following two supersymmetry projectors
\begin{equation}
P_1\epsilon=P_2\epsilon=0\,,
\end{equation}
where the projectors $P_1$ and $P_2$ are given by\footnote{We use the following metric on the tangent space $\eta_{\underline{t}\underline{t}}=0$, $\eta_{\underline{t}\underline{u}}=1$, $\eta_{\underline{u}\underline{u}}=1$ with the rest of the metric diagonal, i.e. $\eta_{\underline{x}\underline{x}}=1$ etc. Further we use $\epsilon_{A_1\ldots A_{10}}=e^{\underline{A}_1}_{A_1}\ldots e^{\underline{A}_{10}}_{A_{10}}e_{\underline{A}_1\ldots\underline{A}_{10}}$ where $e_{\underline{A}_1\ldots\underline{A}_{10}}$ is the tangent space Levi-Civit\`a symbol satisfying $e_{\underline{t}\underline{u}\underline{x}\underline{y}\underline{R}\underline{\theta}_1\ldots\underline{\theta}_4\underline{\varphi}}=+1$. The chirality matrix is $\gamma_{11}=\gamma_{\underline{t}\underline{u}\underline{x}\underline{y}\underline{R}\underline{\theta}_1\ldots\underline{\theta}_4\underline{\varphi}}$ and $\epsilon$ satisfies $\gamma_{11}\epsilon=-\epsilon$.}
\begin{eqnarray}
P_1 & = & \frac{1}{2}\left(1-\gamma_{\underline{t}\underline{u}}\right)\,,\\
P_2 & = & \frac{1}{2}\left(1-i\gamma_{\underline{x}\underline{y}}\right)\,.
\end{eqnarray}
With two projectors the number of preserved supersymmetries is $\text{Tr}1-\text{Tr}P_1-\text{Tr}P_2+\text{Tr}(P_1P_2)$. From this we find that the solution is 1/4 BPS.  The $P_1$ projector is equivalent to demanding $\gamma_{\underline{t}}\epsilon=0$ (the 1/2 BPS axion wave projector) and $P_1$ and $P_2$ together imply the D3-brane projector $\frac{1}{2}\left(1+i\gamma_{\underline{t}\underline{u}\underline{x}\underline{y}}\right)$. The projectors $P_1$ and $P_2$ are in agreement with the supersymmetry analysis of \cite{Donos:2010tu}.

The doubling of the supersymmetries that occurs in the near horizon geometry of the extremal D3-brane does not occur in the case where the extremal D3-brane has a plane wave superposed on it. This can be explicitly verified by substituting the near horizon solution into the above two integrability conditions. Doing so leads again to the projectors $P_1$ and $P_2$. Even more explicitly one can solve the Killing spinor equations and confirm that the only nonzero solutions are 
\begin{equation}
\epsilon=e^{\tfrac{i}{4}g_sku}r^{-1}\eta\,,
\end{equation}
where $\eta$ satisfies $P_1\eta=P_2\eta=0$ and which depends on the 5-sphere coordinates. For explicit expressions for Killing spinors on spheres see for example \cite{Lu:1998nu}.

When the $u$ coordinate is compact we need to impose boundary conditions on the fields. For the axion the natural choice is to demand that it transforms into itself up to a symmetry of the theory. In order for the axion to remain single-valued we must take integer shifts and divide type IIB out by these discrete transformations. The Killing spinor has a dependence on the $u$ coordinate and so we can likewise demand that when going once around the compact circle $\epsilon$ transforms into itself (or possibly minus itself since it is a spinor) up to a symmetry of the theory. The spinors of type IIB supergravity do not transform under the axion shift symmetry and so all we can ask for is that $\epsilon$ transforms into itself up to a sign. This means that we need to impose
\begin{eqnarray}
g_skL & = & 2n\qquad \text{(anti-periodic boundary conditions)}\,,\\
g_skL & = & 4n\qquad \text{(periodic boundary conditions)}\,,
\end{eqnarray}
where $n\in\mathbb{Z}$. At the same time in order for the axion to transform by an integer amount when going once around the $u$ circle we need $2\pi Lk=m\in\mathbb{Z}$. It follows that supersymmetry is preserved if
\begin{eqnarray}
g_s & = & 4\pi\frac{n}{m}\qquad \text{(anti-periodic boundary conditions)}\,,\\
g_s & = & 8\pi\frac{n}{m}\qquad \text{(periodic boundary conditions)}\,.
\end{eqnarray}
This means that for generic values of the string coupling supersymmetry will be broken once $u$ is compactified.

Consider reducing the type IIB theory from 10 to 9 dimensions. Performing a Scherk--Schwarz reduction of the axion shift symmetry requires the supersymmetry parameter $\epsilon$  to be independent of the circle coordinate $u$ \cite{Bergshoeff:2002nv}. This is because the $u$ dependence of the IIB fields is dictated by their transformation properties under the symmetries of type IIB supergravity and under the condition that the lower dimensional equations of motion do not depend on $u$. If we reduce the type IIB Killing spinor equations we obtain the Killing spinor equations of maximal 9-dimensional supergravity. Therefore, 4-dimensional Lifshitz times the 5-sphere is a non-supersymmetric solution of 9-dimensional maximal supergravity (mass deformed by $k$) for if it were a supersymmetric solution it would have to uplift to 10 dimensions to a supersymmetric solution with a $u$-independent Killing spinor and this is not possible.

The analysis of sections \ref{sec:D3branesandwaves} and \ref{sec:susy} can be generalized as follows. In type IIB supergravity one can define besides the RR axion, two other axionic scalars \cite{Bergshoeff:2007aa}; there is one axionic scalar for each Killing vector of the scalar coset manifold. Therefore one can use one of the other two axions and perform a similar analysis to what was done in sections \ref{sec:D3branesandwaves} and \ref{sec:susy}. In the notation of \cite{Bergshoeff:2007aa} the scalar coset manifold is parametrized in terms of the scalars $T$ and $\chi'$, where $\chi'$ is an axion. The solutions then take the form $T=\text{cst}$ and $\chi'$ proportional to $u$. These are again  1/4 BPS intersections of a D3-brane with a wave whose near horizon geometry is a 5-dimensional $z=0$ Schr\"odinger space-time. 

\section{Black deformations}

In this section we will use the D3-brane/axion wave picture to find 4-dimen- sional analytic $z=2$ Lifshitz black brane solutions. The basic observation is the following. If, before we turn on the axion, we have a metric of the form
\begin{eqnarray}
ds^2_{(10)} & = & H^{-1/2}\left(2A_1du dt+A_2du^2+dx^2+dy^2\right)\nonumber\\
&&+H^{1/2}\left(\frac{dR^2}{F}+R^2d\Omega_5^2\right)\,,\label{eq:deformation}
\end{eqnarray}
with $H=l^4R^{-4}$, i.e. in the near horizon limit of the D3-brane, then the effect of adding the axion wave is to deform the function $A_2$ while leaving the functions $A_1$ and $F$ unmodified. Metrics of this form have an everywhere null Killing vector which is $\partial_t$. We will take the function $A_2$ to be proportional to $k$.

For $k=0$, in the near horizon limit of the D3-brane, the 5-dimensional metric described by $t,u,x,y,R$ will be an asymptotically AdS (AAdS$_5$) black string solution with the world-volume of the string described by $t$ and $u$. If we define the coordinates $T$ and $z$ by $\sqrt{2}u=T+z$ and $\sqrt{2}t=-T+z$ then the string stretches in the $z$ direction and is infinitely long. The transverse space, parametrized by $R,x,y$, has translational and rotational symmetries in the $x$ and $y$ directions. The effect of turning on the axion is to deform the asymptotics of the space-time from an AAdS$_5$ metric to an asymptotically $z=0$ Schr\"odinger metric. It also breaks the string's world-volume boost symmetry\footnote{The setup here is reminiscent of that of \cite{Garfinkle:1992zj,Horowitz:1996th} which discusses waves traveling along extremal asymptotically flat black string solutions.}. If we then compactify the direction parametrized by $u$, and reduce along it we obtain a
  4-dimensional black hole solution whose transverse space has translational and rotational symmetries in the $x$ and $y$ directions that is asymptotic to a $z=2$ Lifshitz space-time. In other words a $z=2$ Lifshitz black brane solution.

One may wonder if it is possible to deform the usual black hole deformation of the D3-brane by the axion wave, i.e. to start with the metric
\begin{eqnarray}
ds^2_{(10)} & = & H^{-1/2}\left(-FdT^2+dz^2+dx^2+dy^2\right)\nonumber\\
&&+H^{1/2}\left(\frac{dR^2}{F}+R^2d\Omega_5^2\right)\,,
\end{eqnarray}
where $F=1-MR^{-4}$ (and $H=l^4R^{-4}$ working again only in the near horizon limit) and to deform this by an axion that is proportional to $T+z$. The difficulty one faces is that the space-time now only has asymptotic null Killing vectors while the generator of the horizon is only null at the horizon and timelike everywhere outside. Deforming this space-time such that at infinity we have a wave traveling in the (null) $u$ direction, with $\sqrt{2}u=T+z$, and such that we preserve translational symmetries in the $T,z,x,y$ directions as well as rotational symmetries in the $(x,y)$ plane is expected to be a more difficult task then the black string approach discussed above\footnote{For more details about the problem of combining black holes and plane waves we refer to \cite{LeWitt:2008zx,LeWitt:2009qx}.}. If one were able to solve this problem then upon reduction along $u$ one would obtain $z=2$ Lifshitz black brane solutions. This gives some physical understanding as to why it is difficult to construct generic Lifshitz black brane solutions.

We will now first discuss the black string deformation of the D3-brane and then deform it further by the axion wave. Keeping the matter fields undeformed, i.e. $\chi=ku$, $F_{a_1\ldots a_5}=l^4\omega_{a_1\ldots a_5}$ with the remaining components of $F_5$ determined by self-duality and with $\phi$ some function of $R$, the most general metric that is of the form \eqref{eq:deformation} that is a solution to the type IIB equations of motion (and thus also to the equations of motion of \eqref{eq:5Daction}), assuming first that $k=0$, is given by
\begin{eqnarray}
A_1 & = & F^{1/2}\,,\\
A_2 & = & c_1A_1+c_2A_1\log A_1\,,\\
F & = & 1-MR^{-4}\,,
\end{eqnarray}
with the dilaton given by
\begin{equation}\label{eq:dilaton}
e^\phi=g_sF^{\pm 1/2}\,.
\end{equation}
The metric has a curvature singularity at $R=M^{1/4}$. This however does not imply that the 4-dimensional solution obtained after reduction must also have a curvature singularity at the same value of $R=M^{1/4}$. The constant $c_1$ can always be set to zero by the coordinate transformation $t\rightarrow t-c_1u/2$. We will take $c_2=0$ so that $A_2$ will be proportional to $k$. The two possible signs in the solution for the dilaton is a manifestation of the S-duality invariance (for $\chi=0$) of the theory.

When $k\neq 0$ the solution remains the same except that now $A_2$ is
\begin{equation}\label{eq:A2minus}
A_2=-\frac{g_s^2k^2l^4}{4R^2}+\frac{g_s^2k^2l^4}{2M^{1/2}}A_1\arcsin\frac{M^{1/2}}{R^2}\,,
\end{equation}
when we take the minus sign in the expression \eqref{eq:dilaton} for the dilaton and 
\begin{eqnarray}
A_2 & = & \frac{g_s^2k^2l^4}{4M^{1/2}}A_1\left[\left(1-\log 2\right)\arcsin\frac{M^{1/2}}{R^2}-\left(\arcsin\frac{M^{1/2}}{R^2}\right)\log A_1\right.\nonumber\\
&&\left.+\frac{1}{2}\text{Cl}_2\left(2\arcsin A_1\right)\right]\,,\label{eq:A2plus}
\end{eqnarray}
when we take the plus sign in \eqref{eq:dilaton}. In the expression $\eqref{eq:A2plus}$ the function $\text{Cl}_2(x)$ is the Clausen function that is defined by
\begin{equation}
\text{Cl}_2(x)=-\int_0^x\log\left(2\sin\frac{t}{2}\right)dt\,.
\end{equation}
Since the axion is nonzero whenever $k\neq 0$ the two possible sign choices for the dilaton are no longer related by S-duality, so that we get different solutions for each. These two functions $A_2$ have the property that for $M=0$ they reduce to $g_s^2k^2l^4R^{-2}/4$ which is the near horizon limit of \eqref{eq:functionG} while they vanish for $k=0$. The expression for $A_2$ in \eqref{eq:A2plus} has been written such that each term is positive for $M^{1/4}<R<\infty$. To take the $M=0$ limit of \eqref{eq:A2plus} it useful to use the following two results for Clausen functions \cite{AbramowitzStegun}
\begin{eqnarray}
\text{Cl}_2(\pi-x) & = & \text{Cl}_2(x)-\frac{1}{2}\text{Cl}_2(2x)\,,\\
\text{Cl}_2(x) & = & x-x\log x+\mathcal{O}(x^2)\,,
\end{eqnarray}
with $x=2\arcsin\left(\sqrt{1-A_1^2}\right)=\pi-2\arcsin A_1$.

We next compactify the $u$ direction and reduce to 4 dimensions. To do so we write the metric \eqref{eq:deformation} in the form of a Kaluza--Klein reduction Ansatz as
\begin{equation}
ds_{(10)}^2=e^{-\Phi}ds^2+e^{2\Phi}\left(du+A\right)^2+l^2d\Omega^2_5\,,
\end{equation}
where
\begin{eqnarray}
ds^2 & = & l^3\frac{1}{r}A_2^{1/2}\left[-\frac{1}{r^2 A_2}Fdt^2+\frac{1}{r^2}\left(dx^2+dy^2\right)+\frac{dr^2}{r^2F}\right]\,,\label{eq:ALifmetric}\\
A & = & \frac{F^{1/2}}{A_2}dt\,,\\
e^{2\Phi} & = & l^2\frac{1}{r^2}A_2\,,\label{eq:ALifPhi}
\end{eqnarray}
with $r=l^2R^{-1}$. The asymptotic form is given by \eqref{eq:Lifsolution1} to \eqref{eq:Lifsolution3}. The asymptotic $z=2$ Lifshitz solution \eqref{eq:ALifmetric} to \eqref{eq:ALifPhi} and \eqref{eq:dilaton} (as well as with the 4-dimensional axion constant) thus obtained solves the equations of motion of the 4-dimensional action \eqref{eq:IIB4Daction}. The 4-dimensional result is particularly sensitive to the zeros of $A_2$. In fact these points form curvature singularities. The compactification circle becomes lightlike for points where $A_2=0$. Ideally we would like this to happen inside the region bounded by the horizon. Furthermore, after the reduction $A_2$ cannot be negative. The solution \eqref{eq:A2minus} has one zero in between the horizon and infinity and therefore possesses a naked singularity. The second solution \eqref{eq:A2plus} has one zero that coincides with the horizon, so that the curvature singularity coincides with the horizon.

The type IIB string coupling goes to zero at the horizon in the case where $A_2$ is given by \eqref{eq:A2plus}. Clearly, the supergravity approximation breaks down near the curvature singularity, i.e. before we get to the horizon due to the large curvature. Therefore in the region near the horizon $\alpha'$ corrections to type IIB supergravity will be important.

Even though there is definitely room for improvement, especially to find examples where the curvature singularity is strictly behind the event horizon\footnote{In the last paragraph of section \ref{sec:susy} we mentioned that the solution \eqref{eq:D3braneplanewave1} to \eqref{eq:D3braneplanewave4} can be generalized by considering other coordinate systems on the scalar coset manifold. Here we just mention that these solutions can be black deformed in much the same way as was done above. None of the solutions obtained in this way have the curvature singularity behind the event horizon.} we believe that these ideas provide a complementary view on the more numerical approaches taken in \cite{Danielsson:2009gi,Bertoldi:2009vn,Bertoldi:2009dt,Mann:2009yx}. From a 4-dimensional point of view the problem to find Lifshitz black brane solutions has to do with the fact that it is rather difficult to decouple the differential equations. In our model \eqref{eq:IIB4Daction} the 4-dimensional equations can be uplifted to 5 dimensions and then using the Ansatz
\begin{equation}
d\hat s^2 = \frac{R^2}{l^2}\left(2A_1du dt+A_2du^2+dx^2+dy^2\right)+\frac{l^2}{R^2}\frac{dR^2}{F}
\end{equation}
for the 5-dimensional metric allows one to decouple these differential equations. It would be interesting to study this further from the 4-dimensional point of view in a wider class of models and to see if similar decouplings can be obtained.

\section{Discussion}

The D3-brane/axion wave interpretation of the $z=2$ Lifshitz space-time may be of interest for a number of open problems. For example it has been noted in \cite{Kachru:2008yh} that Lifshitz space-times suffer from divergent tidal forces. This issue has been further elaborated on in \cite{Copsey:2010ya}. These divergent tidal forces persist to exist in the 5-dimensional z=0 Schr\"odinger uplift of the 4-dimensional $z=2$ Lifshitz space-time. It would be interesting to understand better, if only qualitatively, what the right attitude towards this singularity is. We may now have a simple starting point to address this issue. For example it would be interesting to study the effect of introducing the axion wave from the point of view of the world-volume theory of the D3-branes. This will be some deformation of $\mathcal{N}=4$ SYM which may well have an infrared singularity corresponding to the singularity of the $z=0$ Schr\"odinger space-time.

Another potential application of this way of embedding Lifshitz into string theory concerns the problem of holographic renormalization. The 5-dimensional $z=0$ Schr\"odinger space-time is asymptotically AdS$_5$ \cite{Costa:2010cn}. Therefore we can apply the standard holographic renormalization techniques for AAdS$_5$ space-times to construct the counterterm action for the 5-dimen- sional AdS-gravity-axion-dilaton system. Part of this analysis, where it concerns the conformal anomaly, has been done in \cite{Nojiri:1999jj}. Once the complete counterterm action is known for the 5-dimensional AdS-gravity-axion-dilaton system we may be able to obtain the $z=2$ Lifshitz counterterms by dimensional reduction. It would be interesting to compare such an analysis with existing proposals for the holographic renormalization of asymptotically Lifshitz space-times, see e.g. \cite{Ross:2009ar,Balasubramanian:2009rx}.

As discussed at the end of section \ref{sec:D3branesandwaves} (based on \cite{Costa:2010cn}) the Scherk--Schwarz reduction of the bulk $z=0$ Schr\"odinger space-time to the $z=2$ Lifshitz space-time is expected to correspond on the boundary theory to some DLCQ of the field theory dual of the $z=0$ Schr\"odinger space-time. It would be interesting to study the Scherk--Schwarz reduction of a 5-dimensional $z=0$ Schr\"odinger space-time in more detail in particular in relation to the bulk-boundary correspondence.

Regarding the construction of analytic Lifshitz black brane solutions we would like to mention that the approach taken here is not expected to be limited to the particular theory that we work with, but to apply more generally. The general recipe that this approach suggests is to take some 5-dimensional supergravity theory that possesses AdS solutions and that contains an axion and then to look for AAdS$_5$ black string solutions with a null Killing vector on its world-volume and to deform them by an axion wave. The axion profile must be that one which in the case of pure AdS$_5$ deforms it to a $z=0$ Schr\"odinger space-time. Reducing this to 4 dimensions along the direction in which the wave is traveling should then provide us with $z=2$ Lifshitz black brane solutions of various theories. It would be interesting to see if this method can be used to construct examples where the curvature singularity is strictly inside the region bounded by the horizon. This would be highly advantageous as black brane solutions in a holographic context are often the background on which some field theory is defined. Having analytic control of such a background is without doubt desirable. Finally, it would be interesting to see if this method can be used to learn more about the attractor mechanism with Lifshitz scaling (see \cite{Kachru:2011ps} and references therein) where it concerns asymptotically Lifshitz space-times.

\section*{Acknowledgements}

We would like to thank Matthias Blau, Thomas Van Riet and Omid Saremi for useful discussions. The work of J.H. has been supported by the Swiss National Science Foundation and the ``Innovations- und Kooperationsprojekt C-13'' of the Schweizerische Universit\"atskonferenz SUK/CUS. The work of W.C is supported in part by the FWO - Vlaanderen, Project No.
G.0651.11, and in part by the Federal Office for Scientic, Technical and Cultural Affairs through the "Interuniversity Attraction Poles Programme
- Belgian Science Policy" P6$/$11-P.

\end{document}